\documentclass[useAMS,usenatbib]{mn2e}

\usepackage{amsmath}

\usepackage{epsfig}
\usepackage{epstopdf}
\usepackage{lscape} 
\usepackage{natbib}
\usepackage{tabularx}
\usepackage{multirow}
\usepackage{amssymb}
\usepackage{gensymb}

\def\gtrsim{\mathrel{\hbox{\rlap{\hbox{\lower4pt\hbox{$\sim$}}}\hbox{$>$}}}}

\title{FRB repetition and non-Poissonian statistics}
\author[Connor et al.]{
Liam Connor$^{1,2,3}$\thanks{E-mail:\ connor@astro.utoronto.ca}
Ue-Li Pen$^{1, 6, 7}$\thanks{E-mail:\ pen@cita.utoronto.ca}
Niels Oppermann$^{1}$\thanks{E-mail:\ niels@cita.utoronto.ca}
\\
$^1$ Canadian Institute for Theoretical Astrophysics, University of Toronto, M5S 3H8 Ontario, Canada
\\
$^2$ Department of Astronomy and Astrophysics, University of Toronto, 
M5S 3H8 Ontario, Canada
\\
$^3$ Dunlap Institute for Astronomy and Astrophysics, University of Toronto,
Toronto, ON M5S 3H4, Canada
\\
$^6$ Canadian Institute for Advanced Research, Program in Cosmology
and Gravitation
\\
$^7$ Perimeter Institute for Theoretical Physics, 31 Caroline St. N., Waterloo, ON, N2L 2Y5, Canada
}

\begin{document}
\date{\today}
\pagerange{\pageref{firstpage}--\pageref{lastpage}} 
\pubyear{2015}
\maketitle
\label{firstpage}

\begin{abstract}
We discuss some of the claims that have been made regarding the statistics of 
fast radio bursts (FRBs). In an earlier paper \citep{2015arXiv150505535C} 
we conjectured that flicker noise associated 
with FRB repetition could show up in non-cataclysmic neutron star emission models,
like supergiant pulses. We show how the current limits of repetition 
would be significantly weakened if their repeat rate really were non-Poissonian
and had a pink or red spectrum.
Repetition and its statistics have implications for observing strategy, generally favouring 
shallow wide-field surveys, since in the non-repeating scenario survey depth is unimportant. 
We also discuss the statistics of the apparent latitudinal dependence of FRBs, and offer
a simple method for calculating the significance of this effect.  We provide a generalized Bayesian framework for addressing 
this problem, which allows for direct model comparison. 
It is shown how 
the evidence for a steep latitudinal gradient of the FRB rate is less strong than initially suggested 
and simple explanations like increased scattering and sky temperature 
in the plane are sufficient to decrease the low-latitude burst rate, given current data.
The reported dearth of bursts near the plane is further complicated if FRBs have 
non-Poissonian repetition, since in that case the event rate inferred from observation
depends on observing strategy.
\end{abstract}

\newcommand{\be}{\begin{eqnarray}}
\newcommand{\ee}{\end{eqnarray}}
\newcommand{\beq}{\begin{equation}}
\newcommand{\eeq}{\end{equation}}

\section{Introduction}
There is mounting evidence that the new class of transients 
known as fast radio bursts (FRBs) are of extraterrestrial origin.
The most striking features of FRBs are their large dispersion
measures (DMs) -- too high to be attributed to our own Galaxy's
interstellar medium (ISM) --
and their event rate (10$^3$-$10^4$ sky$^{-1}$ day$^{-1}$). They
last for about a millisecond with peak flux of roughly a Jy, and none
has been conclusively shown to repeat. This has led to the 
interpretation that FRBs are cosmological,
since the intergalactic medium (IGM) would naturally provide
DMs between 300-1600 pc cm$^{-3}$ for sources at 
$z\sim0.3$-$1$ \citep{2013Sci...341...53T}. 

Given their apparent phenomenological richness
(polarization, scattering, etc.) and considering
how little we know about their location and physical 
origin, it is likely that FRBs will be of interest to the community for 
years to come, assuming they are not terrestrial.
Though we are in the regime of only a dozen published FRBs, 
at present the conventional wisdom is that they are likely cosmological
in origin \citep{2013Sci...341...53T}, they seem to not repeat regularly
\citep{2015MNRAS.454..457P}, and there is a dearth of bursts 
at low Galactic latitudes \citep{2014ApJ...789L..26P, 2014ApJ...792...19B, 2015MNRAS.451.3278M}.

Based on this premise there have been 
a number of models proposed to describe cataclysmic, cosmological 
FRBs \citep{2012ApJ...760...64M, 2013PASJ...65L..12T, 2014A&A...562A.137F, 2015ApJ...814L..20M}.
However since the field is still in its infancy it is important to
leave as many conceptual doors open as possible;
assumptions about the statistics of the event rate, 
spatial distribution, and repetition are important for the design
and observing strategy of upcoming surveys. In Section \ref{repeat} we explore the 
consequences of repeating FRBs in the case where their 
burst rate is non-Possionian and exhibits a 1/$f$ power spectrum, 
or pink noise. We also investigate the claims of \cite{2015arXiv150701002M}
that FRB 140514 could have been the same source as FRB 110220, 
and comment on its implications. In Section \ref{rate} we
discuss the impact of FRB repetition on survey strategy. In Section \ref{latitude}
we discuss the statistical treatment of the apparent latitudinal dependence 
of the event rate and the lack of detections at low Galactic latitude.


\section{Repeat rates}
\label{repeat}

Though no source has been shown with certainty to repeat, 
the limits on repeatability of FRBs are still weak. Several 
models generically predict repetition, whether periodic or stochastic. 
Galactic flaring stars \citep{2015arXiv150701002M}, radio-bursting 
magnetars \citep{2007arXiv0710.2006P, 2015ApJ...807..179P}, and pulsar planet systems 
\citep{2014A&A...569A..86M} all predict repetition with varying 
rates and burst distributions. 

In \cite{2015arXiv150505535C} it was 
suggested that supergiant pulses from very young pulsars in supernova 
remnants of nearby galaxies could explain the high DMs, Faraday rotation, 
scintillation, and polarization properties of the observed FRBs. We  
proposed that if the repetition of supergiant pulses were non-Poissonian 
(with a pink or red distribution) 
then one might expect several bursts in a short period of time. 
It is also worth mentioning that the statistics and repeat rates of FRBs 
could vary from source to source -- even if 
they come from a single class of progenitors -- so a long follow-up on an individual burst 
may not provide global constraints.
In this 
letter we will refer to stationary Poisson processes (expectation value, $\mu(t)$, is constant in time)
as ``Poissonian". When we discuss non-Poissonian statistics we will be
focusing on stochastic processes that are correlated on varying timescales. For
example we will not discuss periodic signals, which are not Poissonian but 
have already been studied \citep{2015MNRAS.454..457P}.

\subsection{Flicker noise}
\label{flicker}

Pink noise is ubiquitous in physical systems, showing up in 
geology and meteorology, a number 
of astrophysical sources including quasars and the sun, 
human biology, nearly all electronic devices, finance, 
and even music and speech
\citep{1978ComAp...7..103P, 1975Natur.258..317V}.
Though there is no agreed-upon
mathematical explanation for this phenomenon
 \citep{2002physics...4033M}, fluctuations 
are empirically known to be inversely proportional to frequency 
for a variety of dynamical systems. This can be written as 

\begin{equation}
S(f) = \frac{C}{f^\gamma} \,\,\,\,\,\ \textup{if} \,\,\,\,\, f_{\rm{min}} \le f \le f_{\rm{max}}, 
\end{equation}

\noindent where $S(f)$ is the spectral density (i.e. power spectrum), $\gamma$ is 
typically between 0.5-2,
and $f_{\rm{min}}$ and $f_{\rm{max}}$ are cutoffs beyond which the power law 
does not hold. In this paper will describe these distributions as having flicker noise. 

In the case of a time-domain astronomical 
source, this results in uniformity on short timescales, i.e. a burst of 
clustered events followed by extended periods of quiescence. 
If FRBs were to exhibit such flicker noise then their repetition would 
not only be non-periodic, but would also have a time-varying 
pulse rate and, more importantly, variance. Therefore the number of events
seen in a follow-up observation would depend strongly on 
the time passed since the initial event.

In \cite{2015MNRAS.454..457P}
the fields of eight FRBs discovered between 2009 and 2013
were followed up from April to October of 2014, for an average 
of 11.4 hours per field. During this follow-up programme 
FRB 140514 was found in the same beam as FRB 110220, however
the authors argue that it is likely a new source due to its lower DM. 
After its discovery, the field of 140514 was monitored 
five more times, starting 41 days later on 2014-06-24, without seeing anything.
Under the assumption that 140514 was a new FRB 
that only showed up in the same field 
coincidentally and that the repeat rate is constant,
 \cite{2015MNRAS.454..457P} rule out repetition with a period $P \le$ 8.6 
hours and reject 8.6 $<$ $P$ $<$ 21 hours with 90$\%$ confidence.
However it is possible that one or both of those premises 
is invalid, so it is useful to explore the possibility of non-stationary 
repeat rate statistics and repeating FRBs with variable DM. 

If the statistics of the FRB's repeat rate were
non-Poissonian and initial bursts from FRBs were to have aftershocks
similar to earthquakes, then the non-immediate follow-up observations 
impose far weaker repeat rate limits than has been suggested. We 
constructed a mock follow-up observation of the eight FRBs whose
fields were observed in \cite{2015MNRAS.454..457P}. We then 
asked how many bursts are seen to repeat if we do an immediate
follow-up vs. a follow-up several years after the initial event 
at times corresponding 
to the actual observations carried out. 

We run a simple Monte Carlo 
simulation with one sample per hour and a probability of 0.5 
that a given sample has a burst in it. The repeat rate of once per two hours
is chosen arbitrarily and should not affect the comparison. To get the 
1$/f^{\gamma}$ distribution we
take an uncorrelated Gaussian time stream centred on 0 and move to Fourier space, then 
multiply by $f^{\gamma / 2}$, which gives a power spectrum with the desired shape. 
We then inverse Fourier transform back to get the pink or red time stream.
We then take samples with a positive value to contain a pulse and samples with a negative value to contain none.. 
In the stationary Poisson case, the rate of bursts in 
the immediate follow-up is the same as the multi-year follow-up 
since all times are 
statistically equivalent. However with flicker noise 
the variance is strongly time-dependent. If we imagine an object that 
repeated on average once per two hours, then if those pulses were Poisson-distributed 
the probability of seeing zero bursts in 11.4 hours or longer is $\sim0.007$. With 
pink noise one expects this roughly 20$\%$ of the time, since the system 
prefers either to be in ``on" or ``off" mode. If the average repeat period were more 
like 5-20 hours, then we would often see nothing in a multi-day follow-up 
observation that took place weeks or years after the initial event.

\begin{figure}
  \centering
   \includegraphics[trim={1in, 0in, 1in, 0in}, width=0.45\textwidth, height=0.29\textwidth]{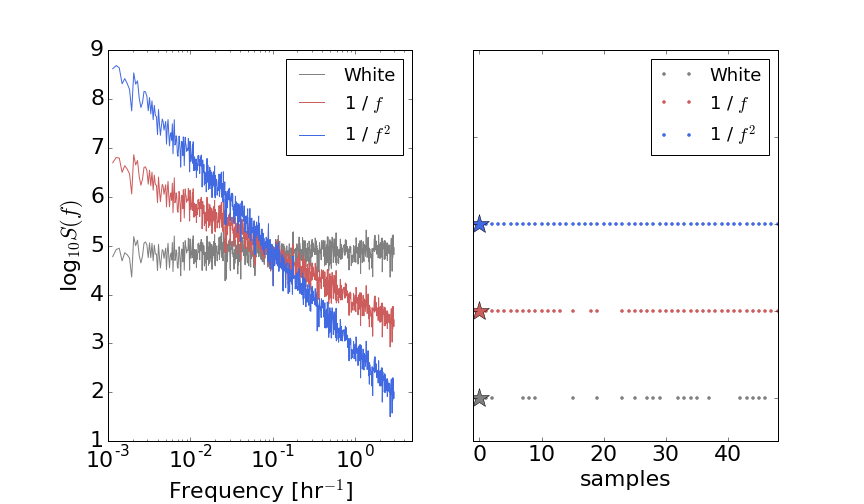}
   \caption{Realization of our mock follow-up Monte Carlo. \textit{Left panel}: Power spectrum for pulse arrival times of a single FRB.
   Grey shows a flat spectrum, corresponding to the often assumed Poissonian 
   repetition rate. The red and blue spectra show flicker noise, with pink (1/$f$) 
   noise and Brownian (1/$f^2$) noise respectively. \textit{Right panel}: 
   We found the first ``event" in our Monte Carlo (represented by a star) for the three different spectra 
   and plotted their behaviour in the subsequent 48 hours of follow-up. Though the average
   probability over the whole simulation is 0.5 for each distribution, when we zoom in 
   on this short period the strong time-like correlations 
   in the $1/f^\gamma$ cases means there are many repetitions: they are in an ``on" state at this time.}
   \label{FIG-RATE}
\end{figure}

This is consistent with what \cite{2015MNRAS.454..457P} saw, though the 
conclusions differ depending on the assumed statistics. In Figure \ref{FIG-RATE}
we show a sample from this simulation for three repeat distributions. The right panel 
shows how, if an FRB's burst rate has long-term correlations (1$/f^\gamma$), 
the likelihood of a repeat is greatly increased if the follow-up observation is 
immediately after the initial event, rather than months or years after.

\subsection{FRBs 110220 and 140514}
Using the event rate of roughly $10^4$ sky$^{-1}$ day$^{-1}$
from \cite{2013Sci...341...53T}, it 
was originally reported that the probability of seeing a 
new FRB in the field of 110220 during the 85 hours of follow-up 
was $0.32$  
\citep{2015MNRAS.447..246P}. It was then pointed out by \cite{2015arXiv150701002M} 
that this underestimated the coincidence by an order of magnitude, 
since they estimated the rate in any one of the 13 beams, 
while the new event occurred in the identical beam.
The probability also dropped due to the updated daily event rate,
given the \cite{2013Sci...341...53T} estimate is now thought likely to be too high. 
In general we expect the true rate of FRBs to be lower than what is 
reported due to non-publication bias: If archival data are searched and 
nothing is found, it is less likely to be published than if something is found. 
That said, using the rate calculated by \cite{2015arXiv150500834R} and following
the procedure of \cite{2015arXiv150701002M}, we find the likelihood of finding a new burst to be 
between 0.25-2.5$\%$.

\begin{figure}
  \centering
   \includegraphics[trim={0in, 0in, 0in, 0in}, width=0.43\textwidth, height=0.3\textwidth]{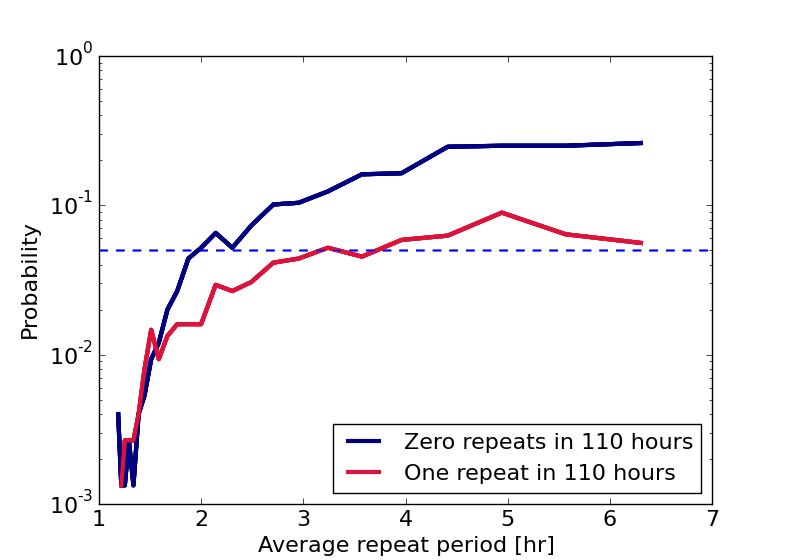}
   \caption{The probability of seeing zero (blue curve) or one (red curve) repeat burst
   in 110 hours of follow-up, assuming a 1/$f$ distribution.  
   The curves are derived from a simple Monte Carlo in which a
   pink distribution was generated with one sample per hour, 
   and we asked how many bursts were seen in 110 samples. Thousands of 110-hour realizations
   were then averaged for each repeat period, $P$. 
   Even with an average 
   repeat rate of once every two hours, there is still a 5$\%$ chance (indicated by the dashed line) of seeing one or
   fewer bursts, despite the expected value of 55.}
   \label{FIG-hist}
\end{figure}

Given the relatively low probability of finding 
a new FRB in the same field and since there are models that predict
burst repetition with variable DMs \citep{2015arXiv150505535C, 2015arXiv150701002M}
one can ask the question: If one FRB out of eight is found to
repeat during 110 hours of follow-up (including extra time spent on 140514), 
what are the limits on the average
repeat period? Another way of asking this question is what is the probability of 
 some number of repetitions during the 110 hours, given a repeat rate. The answer to 
this question depends strongly on the power spectrum's shape. For the sake of example, if the average 
repeat rate is once per two hours, then the probability of one repeat or fewer in the Poisson
case is effectively zero. With a pink distribution it is closer to $5\%$, even though 
the expected number would be 55. This is 
shown in Figure \ref{FIG-hist}, in which we plot the probability of seeing zero or one repeat burst 
(the two options for FRB 140514), given some average repetition period, $P$. 
We generate the pink distribution in the same way described in Section \ref{flicker}, 
using one-hour samples and a long-wavelength cutoff at 1.2 million hours. Though 
it was taken arbitrarily, the probability of seeing no bursts should depend only 
weakly on this cutoff. Since the variance scales logarithmically with this number, 
there is only roughly a factor of three 
difference in total power between our choice and $f_{\rm{min}}\sim$ an inverse Hubble time. 
While we remain agnostic about the relationship between 140514 and 110220, with 
non-Poissonian repetition it is possible to have a relatively high repeat rate 
and to see either one or zero repeat bursts in several days of observation.

\section{Event rates and total number of sources}
\label{rate}

If FRBs were found to repeat, their statistics and the
average frequency of their repetition 
should affect the search strategy of upcoming surveys. 
For instance, if it were found that FRBs repeated,
on average, five times a day, then the number of unique 
sources would be five times smaller than the per-sky 
daily event rate. This means the daily rate 
$3.3^{+5.0}_{-2.5}\times10^3$ sky$^{-1}$ estimated by 
\cite{2015arXiv150500834R} would be produced by
only $\sim$160-1600 sources. In this scenario 
there is no FRB in most pixels on the sky, which means
one could integrate on most patches forever without 
seeing an event. An example of this strategy is the VLA millisecond search, 
in which $\sim40\%$ of the time was spent at a single pointing, and
almost three quarters of the time was spent at just three locations \citep{2015ApJ...807...16L}.
It is possible that pointing-to-pointing event rate variance contributed 
to their not seeing anything.

We therefore warn that deep surveys are at a disadvantage 
to those that sweep large regions of the sky 
(CHIME  \citep{2014SPIE.9145E..22B}, 
UTMOST\footnote{http://www.caastro.org/news/2014-utmost}, HIRAX)
because the non-repeating 
scenario is unaffected; whereas shallow observations 
should not hurt the detection rate, no matter what their repetition. 
Ideally, a survey
needs only to spend a few dispersion delay times on each beam 
before moving on. 

\section{Latitudinal dependence}
\label{latitude}

There is now evidence that the FRB rate is nonuniform on the sky, 
with fewer detectable events at low Galactic latitudes \citep{2014ira..book.....B}.
However the statistical significance of this finding may be 
overestimated. \cite{2014ApJ...789L..26P}
compute the probability of the disparity between the number of 
bursts seen in the high- and intermediate-latitude ($|b| < 15\degree$)
components of the High Time Resolution Universe survey 
(HTRU). They calculate the probability of seeing $N=0$ in the intermediate 
latitude survey and $M=4$ in 
the high-latitude, despite having searched 88$\%$ more data in the former, and they 
rule out the uniform sky hypothesis with 99.5$\%$ certainty. We would point out 
that in general $P(N|M)$ describes a very specific outcome, and it would be 
more appropriate to include all outcomes equally or more unlikely.
That number might also be multiplied by two, since if the survey found 
four intermediate-latitude FRBs and zero high-lat ones, we would ask the same 
question. 

But a simpler approach to this problem would be analogous to a series of 
coin flips. If a coin were flipped four times, the probability of seeing all heads is 1/16, 
or 6.25$\%$. This is a factor of six higher than  the analogous analysis of 
\citet{2014ApJ...789L..26P}.  We can test the null hypothesis that the coin is fair, and using
the binomial statistic would conclude that the outcome is consistent with a 
fair coin at 95\% confidence, differing from the conclusion of \citet{2014ApJ...789L..26P}.
In the FRB case the Universe is flipping a coin each time a new burst 
appears, with some bias factor due to things like different integration times.
HTRU 
has since reported five more bursts in the high Galactic region, but using a dataset 
that spent $\sim$2.5 times more time at high latitudes. Below we try and quantify the likelihood of 
this.

If one wants to 
compare two statistical hypotheses, then the claims of each should 
be treated as true and their likelihood discrepancy should be computed.
In the case of testing the abundance of FRBs at high latitudes,
the sky should be partitioned into high and low regions a priori 
(e.g., the predefined high-latitude HTRU and its complement). The rate in both regions 
is then taken to be the same, and the likelihood of a given spatial distribution of observed
sources can be calculated. This situation is naturally
described by a biased binomial distribution with a fixed number of events. Suppose
a total of $K$ FRBs are observed in a given survey. We can ask the question, what is the probability of 
seeing $M$ events in the high region and ($K-M$) events in the lower region?
This probability can be calculated 
as 

\begin{equation}
\label{eq-binomial}
P(M | \, {K}, p) =  \binom{K}{M} \, p^{M} (1-p)^{K-M} ,
\end{equation}

\noindent where $p$ is the probability that an event happens to show up in the 
high region. In a survey where more time is spent on one part of the 
sky than the other, $p=\alpha/(\alpha+1)$, where $\alpha$ is the ratio of 
time spent in the high-latitude region vs. the intermediate region. In the case of the HTRU 
survey, $K=9$ and since none were found 
in the low-latitude region, $M=9$. Roughly 2500 hours were spent searching the upper region
and $\sim1000$ hours were spent at $|b| < 15\degree$, giving $\alpha=2.5$. Using Equation 
\ref{eq-binomial}, this outcome is only $\sim5\%$ unlikely. 

The problem is given a quasi-Bayesian treatment by
\cite{2014ApJ...789L..26P}, which gives the following.

\begin{equation}
\label{eq-petroff}
P(N | \,M) =  \alpha^{N} (1 + \alpha)^{-(1+M+N)} \frac{(M+N)!}{M!N!}
\end{equation}

\noindent This gives a probability of $\sim$3.5$\%$, using all nine FRBs. This method 
is Bayesian in the sense that they marginalize over the unknown rate 
and calculate a likelihood, but they then calculate a confidence 
and do not look at a posterior.  






The most obvious difference between the approach we have offered (biased coin-flip) 
and the quasi-Bayesian method is that we take $M+N$ to be fixed. It follows to ask 
whether or not we \textit{should} regard the total number of FRBs as ``given"? 
We believe the answer is yes, since this is one of the few quantities that we 
have actually measured, along with $M$ and $N$. What we are really trying 
to infer is how much larger $\mu_{\textup{high}}$ is than $\mu_{\textup{low}}$, so these 
rates should not be marginalized over. 

To consider only the likelihood can give misleading results. 
For example, as more and more FRBs are detected, the 
likelihood of the particular observed values for $N$ and $M$ 
will become smaller and smaller, due to the sheer number of 
possible tuples $(N,M)$.
To decide whether or not there is evidence for FRBs to occur with a 
higher probability at high latitudes, we can instead use the formalism 
of Bayesian model selection. This formalism does not aim to rule out a
particular model, it only compares the validity of two models. 
For this, we formulate two specific models, Model~1 in which we
assume that $p_1=\alpha/(1+\alpha)$ as above (i.e. uniform rate across the sky), and Model~2 in 
which we regard $p$ as a free parameter, equipped with a flat prior 
between 0 and 1. The model selection will then be based on the comparison
of the posterior probabilities for the two models,

\begin{equation}
	\frac{P(\mathrm{Model}~1|M,K)}{P(\mathrm{Model}~2|M,K)} = \frac{P(M|\mathrm{Model}~1,K)}{P(M|\mathrm{Model}~2,K)},
\end{equation}

\noindent where we have assumed equal prior weights for the two models. 
Using the binomial likelihood, Eq.~\eqref{eq-binomial}, and
marginalizing over the unknown probability $p$ in the case of Model~2, 
this ratio is easily calculated to be

\begin{equation}
\frac{P(M|\mathrm{Model}~1,K)}{P(M|\mathrm{Model}~2,K)} = (K + 1) \, \binom{K}{M} \, p_1^M \, \left(1 - p_1\right)^{K-M}.
\end{equation}

\noindent For the observations discussed above with $M = K = 9$ 
and $\alpha = 2.5$, we find a ratio of $0.48$, so there is
no strong preference for either of the two models.

\section{Conclusions}

The search for FRBs with multiple surveys that have disparate sensitivities, 
frequency coverage, and survey strategy (not to mention non-publication bias) 
has made it difficult even to 
estimate a daily sky rate. 
That combined with the relatively low number of total FRBs observed 
has meant that dealing with their statistics can be non-trivial. In 
the case of repetition, we remind the reader that several non-cataclysmic 
models for FRBs are expected to repeat. In the case of supergiant 
pulses from pulsars, SGR radio flares, or even Galactic flare stars, it is possible 
that this repetition would be non-stationary and might exhibit strong correlations 
in time. We have shown that if the repetition had some associated flicker noise 
and its power spectrum were 1/$f^\gamma$, then one should expect the repetition 
rate to be higher immediately after the initial FRB detection. Therefore follow-up 
observations to archival discoveries that take place years or 
months after the first event would not provide strong upper limits. 
This would also mean that if no burst is found in a given beam after some 
integration time, then it is unlikely that one will occur in the following integration, and therefore 
a new pointing should be searched. In other words, shallow fast surveys would be favourable. 

In Section \ref{latitude} we offered a simple way of quantifying the 
latitudinal dependence of FRBs with a binomial distribution. This 
is akin to a biased coin flip, in which we ask ``what is the probability of 
$M$ bursts being found in one region and $N$ bursts in its complement, given 
$\alpha$ times more time was spent in the former". Like \cite{2015arXiv150500834R} we argue
that the jury is still out on the severity of the latitudinal 
dependence. With current data the preference for FRBs to 
be discovered outside of the plane seems consistent with
sky-temperature effects and increased scattering, or even pure chance. 
Whether or not more sophisticated explanations 
(e.g., \citealt{2015MNRAS.451.3278M}) are required remains to be seen. 
We also provided a Bayesian framework for model comparison,
which can be used in the limit where large numbers of FRBs have
been detected. 

\section{Acknowledgements}

We thank NSERC for support.

\newcommand{\araa}{ARA\&A}   
\newcommand{\afz}{Afz}       
\newcommand{\aj}{AJ}         
\newcommand{\azh}{AZh}       
\newcommand{\aaa}{A\&A}      
\newcommand{\aas}{A\&AS}     
\newcommand{\aar}{A\&AR}     
\newcommand{\apj}{ApJ}       
\newcommand{\apjs}{ApJS}     
\newcommand{\apjl}{ApJ}      
\newcommand{\apss}{Ap\&SS}   
\newcommand{\baas}{BAAS}     
\newcommand{\jaa}{JA\&A}     
\newcommand{\mnras}{MNRAS}   
\newcommand{\nat}{Nat}       
\newcommand{\pasj}{PASJ}     
\newcommand{\pasp}{PASP}     
\newcommand{\paspc}{PASPC}   
\newcommand{\qjras}{QJRAS}   
\newcommand{\sci}{Sci}       
\newcommand{\solphys}{Solar Physics}       %
\newcommand{\sova}{SvA}      
\newcommand{\aap}{A\&A}
\newcommand\jcap{{J. Cosmology Astropart. Phys.}}%
\newcommand{\prd}{Phys. Rev. D}

\bibliography{frb_statistics}

\begin{thebibliography}{}

\bibitem[\protect\citeauthoryear{{Bandura}}{{Bandura}}{2014}]{2014SPIE.9145E..22B}
{Bandura} K. e.~a.,  2014, in Society of Photo-Optical Instrumentation
  Engineers (SPIE) Conference Series Vol.~9145 of Society of Photo-Optical
  Instrumentation Engineers (SPIE) Conference Series, {Canadian Hydrogen
  Intensity Mapping Experiment (CHIME) pathfinder}.
p.~22

\bibitem[\protect\citeauthoryear{{Burke} \& {Graham-Smith}}{{Burke} \&
  {Graham-Smith}}{2014}]{2014ira..book.....B}
{Burke} B.~F.,  {Graham-Smith} F.,  2014, {An Introduction to Radio Astronomy}

\bibitem[\protect\citeauthoryear{{Burke-Spolaor} \&
  {Bannister}}{{Burke-Spolaor} \& {Bannister}}{2014}]{2014ApJ...792...19B}
{Burke-Spolaor} S.,  {Bannister} K.~W.,  2014, \apj, 792, 19

\bibitem[\protect\citeauthoryear{{Connor}, {Sievers} \& {Pen}}{{Connor}
  et~al.}{2015}]{2015arXiv150505535C}
{Connor} L.,  {Sievers} J.,    {Pen} U.-L.,  2015, ArXiv e-prints

\bibitem[\protect\citeauthoryear{{Falcke} \& {Rezzolla}}{{Falcke} \&
  {Rezzolla}}{2014}]{2014A&A...562A.137F}
{Falcke} H.,  {Rezzolla} L.,  2014, \aap, 562, A137

\bibitem[\protect\citeauthoryear{{Law}, {Bower}, {Burke-Spolaor}, {Butler},
  {Lawrence}, {Lazio}, {Mattmann}, {Rupen}, {Siemion} \& {VanderWiel}}{{Law}
  et~al.}{2015}]{2015ApJ...807...16L}
{Law} C.~J.,  {Bower} G.~C.,  {Burke-Spolaor} S.,  {Butler} B.,  {Lawrence} E.,
   {Lazio} T.~J.~W.,  {Mattmann} C.~A.,  {Rupen} M.,  {Siemion} A.,
  {VanderWiel} S.,  2015, \apj, 807, 16

\bibitem[\protect\citeauthoryear{{Macquart} \& {Johnston}}{{Macquart} \&
  {Johnston}}{2015}]{2015MNRAS.451.3278M}
{Macquart} J.-P.,  {Johnston} S.,  2015, \mnras, 451, 3278

\bibitem[\protect\citeauthoryear{{Maoz}, {Loeb}, {Shvartzvald}, {Sitek},
  {Engel}, {Kiefer}, {Kiraga}, {Levi}, {Mazeh}, {Pawlak}, {Rich}, {Tal-Or} \&
  {Wyrzykowski}}{{Maoz} et~al.}{2015}]{2015arXiv150701002M}
{Maoz} D.,  {Loeb} A.,  {Shvartzvald} Y.,  {Sitek} M.,  {Engel} M.,  {Kiefer}
  F.,  {Kiraga} M.,  {Levi} A.,  {Mazeh} T.,  {Pawlak} M.,  {Rich} R.~M.,
  {Tal-Or} L.,    {Wyrzykowski} L.,  2015, ArXiv e-prints 1507.01002

\bibitem[\protect\citeauthoryear{{Mickaliger}, {McLaughlin}, {Lorimer},
  {Langston}, {Bilous}, {Kondratiev}, {Lyutikov}, {Ransom} \&
  {Palliyaguru}}{{Mickaliger} et~al.}{2012}]{2012ApJ...760...64M}
{Mickaliger} M.~B.,  {McLaughlin} M.~A.,  {Lorimer} D.~R.,  {Langston} G.~I.,
  {Bilous} A.~V.,  {Kondratiev} V.~I.,  {Lyutikov} M.,  {Ransom} S.~M.,
  {Palliyaguru} N.,  2012, \apj, 760, 64

\bibitem[\protect\citeauthoryear{{Milotti}}{{Milotti}}{2002}]{2002physics...4033M}
{Milotti} E.,  2002, ArXiv Physics e-prints

\bibitem[\protect\citeauthoryear{{Mingarelli}, {Levin} \& {Lazio}}{{Mingarelli}
  et~al.}{2015}]{2015ApJ...814L..20M}
{Mingarelli} C.~M.~F.,  {Levin} J.,    {Lazio} T.~J.~W.,  2015, \apjl, 814, L20

\bibitem[\protect\citeauthoryear{{Mottez} \& {Zarka}}{{Mottez} \&
  {Zarka}}{2014}]{2014A&A...569A..86M}
{Mottez} F.,  {Zarka} P.,  2014, \aap, 569, A86

\bibitem[\protect\citeauthoryear{{Pen} \& {Connor}}{{Pen} \&
  {Connor}}{2015}]{2015ApJ...807..179P}
{Pen} U.-L.,  {Connor} L.,  2015, \apj, 807, 179

\bibitem[\protect\citeauthoryear{{Petroff}, {Bailes}, {Barr}, {Barsdell},
  {Bhat}, {Bian}, {Burke-Spolaor}, {Caleb}, {Champion}, {Chandra} \& {Da
  Costa}}{{Petroff} et~al.}{2015}]{2015MNRAS.447..246P}
{Petroff} E.,  {Bailes} M.,  {Barr} E.~D.,  {Barsdell} B.~R.,  {Bhat} N.~D.~R.,
   {Bian} F.,  {Burke-Spolaor} S.,  {Caleb} M.,  {Champion} D.,  {Chandra} P.,
    {Da Costa} 2015, \mnras, 447, 246

\bibitem[\protect\citeauthoryear{{Petroff}, {Johnston}, {Keane}, {van Straten},
  {Bailes}, {Barr}, {Barsdell}, {Burke-Spolaor}, {Caleb}, {Champion}, {Flynn},
  {Jameson}, {Kramer}, {Ng}, {Possenti} \& {Stappers}}{{Petroff}
  et~al.}{2015}]{2015MNRAS.454..457P}
{Petroff} E.,  {Johnston} S.,  {Keane} E.~F.,  {van Straten} W.,  {Bailes} M.,
  {Barr} E.~D.,  {Barsdell} B.~R.,  {Burke-Spolaor} S.,  {Caleb} M.,
  {Champion} D.~J.,  {Flynn} C.,  {Jameson} A.,  {Kramer} M.,  {Ng} C.,
  {Possenti} A.,    {Stappers} B.~W.,  2015, \mnras, 454, 457

\bibitem[\protect\citeauthoryear{{Petroff}, {van Straten}, {Johnston},
  {Bailes}, {Barr}, {Bates}, {Bhat}, {Burgay}, {Burke-Spolaor} \&
  {Champion}}{{Petroff} et~al.}{2014}]{2014ApJ...789L..26P}
{Petroff} E.,  {van Straten} W.,  {Johnston} S.,  {Bailes} M.,  {Barr} E.~D.,
  {Bates} S.~D.,  {Bhat} N.~D.~R.,  {Burgay} M.,  {Burke-Spolaor} S.,
  {Champion} 2014, \apjl, 789, L26

\bibitem[\protect\citeauthoryear{{Popov} \& {Postnov}}{{Popov} \&
  {Postnov}}{2007}]{2007arXiv0710.2006P}
{Popov} S.~B.,  {Postnov} K.~A.,  2007, ArXiv e-prints

\bibitem[\protect\citeauthoryear{{Press}}{{Press}}{1978}]{1978ComAp...7..103P}
{Press} W.~H.,  1978, Comments on Astrophysics, 7, 103

\bibitem[\protect\citeauthoryear{{Rane}, {Lorimer}, {Bates}, {McMann},
  {McLaughlin} \& {Rajwade}}{{Rane} et~al.}{2015}]{2015arXiv150500834R}
{Rane} A.,  {Lorimer} D.~R.,  {Bates} S.~D.,  {McMann} N.,  {McLaughlin} M.~A.,
     {Rajwade} K.,  2015, ArXiv e-prints 1505.00834

\bibitem[\protect\citeauthoryear{{Thornton}}{{Thornton}}{2013}]{2013Sci...341...53T}
{Thornton} D. e.~a.,  2013, Science, 341, 53

\bibitem[\protect\citeauthoryear{{Totani}}{{Totani}}{2013}]{2013PASJ...65L..12T}
{Totani} T.,  2013, \pasj, 65, L12

\bibitem[\protect\citeauthoryear{{Voss} \& {Clarke}}{{Voss} \&
  {Clarke}}{1975}]{1975Natur.258..317V}
{Voss} R.~F.,  {Clarke} J.,  1975, \nat, 258, 317

\end{thebibliography}
\bibliographystyle{mn2e}

\label{lastpage}

\end{document}